\documentclass{article}
\usepackage{amsmath}
\usepackage{amssymb}

\input{tcilatex}
\begin{document}

\begin{center}
{\LARGE Behaviour of a spin-1/2 particle in Schwarzschild embedded in an
electromagnetic universe }

A. Al-Badawi

Department of Physics, Al-Hussein Bin Talal University, P. O. Box: 20,
71111, Ma'an, Jordan

E-mail: ahmadalbadawi@hotmail.com

{\Large Abstract}
\end{center}

The Dirac equation is considered in Schwarzschild black hole immersed in an
electromagnetic universe with charge coupling. The equations of the charged
spin-1/2 particle is separated into radial and angular equations by adopting
the Newman-Penrose formalism. The angular equations obtained are similar to
the Schwarzschild geometry. For the radial equations we manage to obtain the
one dimensional Schr\"{o}dinger-type wave equations with effective
potentials. Due to the presence of electromagnetic field from the
surroundings, the interaction with the charged spin-1/2 is considered.
Finally, we study the behaviour of the potentials by plotting them as a
function of radial distance and expose the effect of the external parameter,
charge and the frequency of the particle on them.

\bigskip

\section{Introduction}

In general relativity, Schwarzschild (S) black hole is the solution to the
Einstein field equations that describes the gravitational field outside a
spherical mass, while Reissner-Nrdstr\"{o}m (RN) black hole is the solution
to the Einstein-Maxwell field equations that describes the gravitational
field outside a spherical charged mass. Thus the electromagnetic (em) field
in RN\ solution originates from the S mass. The above well known solutions S
and RN, are isolated system. However, the notion of considering the system
in a uniform external em field is a well known subject by now. For example
the S black hole immersed in a homogeneous em field has been introduced in
[1-3]. This represents an external field of non rotating uncharged mass
immersed in external em and gravitational field, Bertotti-Robinson (BR)
universe[4-5]. Mathematically, they interpolate two exact solutions of
Einstein's equations. As a result of this embedding the horizon and particle
geodesics are modified. The extension of an S source in an external em field
has been obtained by Halilsoy et. al. [6]. In the last reference, they
present the metric of a S black hole coupled to an external, stationary em
BR solution. Limiting cases of their metric include the case of a stationary
em universe in which conformal curvature arises due to rotation. Again, the
horizon radius of such a black hole and particle geodesics changed. Later
on, the same authors in [6] have introduced a new metric describes the RN
black hole coupled to an external, stationary em static field [7]. We will
in this paper study the Dirac equation by using the metric obtained in [7].

On the other hand, the behaviour of a spin-1/2 particle in different
backgrounds has been studied by many authors [8-10]. Similarly, the
behaviour of a spin-1/2 particle around a charged black hole is studied by
Mukhopadhyay [11], where he studied the behaviour of the potential by
varying the charge of the black hole. He also study the space-dependent
reflection and transmission coefficients and showed that as the potential
barrier level decreases, the corresponding transmission probability
increases. The separation of variables of the spin-1 and 3/2 field equation
is performed in details in the S geometry by means of the Newman Penrose
(NP) formalism [12,13]. Zecca showed that, as a consequence of the
particular nature of the spin coefficients it is shown, by induction, that
the massive field equations can be separated for arbitrary spin. Recently,
Dirac equation was examined in Kerr-Taub-NUT spacetime [14], by using
Boyer-Lindquist coordinates they obtained \ exact solution of the angular
equations for some special cases and exposed the effect of the NUT parameter
from from the effective potentials plots. Other studies of Dirac equation in
different backgrounds, Nutku helicoid spacetime [15], Kerr--Newman--AdS
black hole geometry [16] and in the background of the Kerr--Newman family,
which has been considered in several studies [17-18].

The spacetime we are considering consists of a central mass plus an external
em radiation which may not be attributed to the charge of the mass [1,7].
The metric that describe this spacetime interpolates the two well-known
solutions of general relativity, the S and the BR solutions. We shall refer
to this solution as the S em black hole (SEBH). Here, we study the solution
of the Dirac equation in SEBH. The set of equations representing the Dirac
equation in the NP formalism is decoupled into a radial (function of
distance $r$ only) and an angular parts (function of angle $\theta $ only).
The solution of the angular equation is given in terms of standard spherical
harmonics as in the S geometry. The radial equations are discussed and the
radial wave equations with effective potentials are obtained. Finally in
order to understand and expose the effect of the external parameter, charge
and the frequency of the spin-1/2 particle on the potentials, curves are
plotted and discussed.

Our paper is organized as follows: in section 2, we present the Dirac
equation in SEBH and separate them into two parts. In section 3, solutions
of the angular and radial equations. In section 4, we study the behaviour of
the potentials by varying the external parameter, the charge and the
frequency by plotting the potentials as a function of radial distance.
Finally we make concluding remarks and digress about future applications of
this research.

\section{Dirac equation in SEBH}

The SEBH metric we are dealing with represents the non-linear superposition
of the S solution and the BR solution[1,7], is given by

\begin{equation}
ds^{2}=\frac{\Delta }{r^{2}}dt^{2}-\frac{r^{2}}{\Delta }dr^{2}-r^{2}\left(
d\theta ^{2}+\sin ^{2}\theta d\phi ^{2}\right) ,
\end{equation}

where 
\begin{equation}
\Delta =r^{2}-2Mr+M^{2}\left( 1-a^{2}\right) .
\end{equation}%
in which, $M$ is the \ S mass coupled to\ an external electromagnetic field
and $a\left( 0<a\leq 1\right) $ is the external parameter. This metric
satisfies all the required limits as boundary conditions:

\begin{eqnarray}
\left( a=1\right) &\rightarrow &\left( the\ S\,Solution\right) \\
\left( 0<a\leq 1\right) &\rightarrow &\left( the\ S+BR\,or\,SEBH\
Solution\right) .  \notag
\end{eqnarray}

The case $\left( a=0\right) ,$ which is excluded, is the extremal RN which
is transformable to BR metric. As a result of the coupling of S and BR, the
horizon shrinks to a significant degree. The horizon is given by%
\begin{equation}
r_{h}=M\left( 1+a\right) .
\end{equation}%
It is clear that $r_{h}\leq 2m$, since $\left( 0<a\leq 1\right) .$ The
vector potential for the above metric is given by 
\begin{equation}
A_{\mu }=\left( \frac{M\sqrt{1-a^{2}}}{2a\left( r-r_{h}\right) }%
,0,0,0\right) .
\end{equation}

Using NP formalism, the Dirac equation can be written as 
\begin{subequations}
\begin{equation}
\sigma _{AB^{\prime }}^{i}\partial _{i}P^{A}+i\mu _{p}\overline{Q}%
_{B^{\prime }}\epsilon _{C^{\prime }B^{\prime }}=0
\end{equation}%
\begin{equation}
\sigma _{AB^{\prime }}^{i}\partial _{i}Q^{A}+i\mu \overline{_{p}P}%
_{B^{\prime }}\epsilon _{C^{\prime }B^{\prime }}=0
\end{equation}

where $P^{A}$ and $Q^{A}$ are the pair of spinors. $\mu _{p}/\sqrt{2}$ is
the mass of the particles and $\sigma _{AB^{\prime }}^{i}$ is the Pauli
matrix.

Introducing the null tetrad \ 1-form of NP formalism, we make the choice of
the following null tetrad basis 1-forms $\left( l,n,m,\overline{m}\right) $
of the NP formalism [19, 20] in terms of this new basis Pauli matrices can
be written as 
\end{subequations}
\begin{equation}
\sigma _{AB^{\prime }}^{i}=\frac{1}{\sqrt{2}}%
\begin{pmatrix}
l^{\mu } & m^{\mu } \\ 
\overline{m}^{\mu } & n^{\mu }%
\end{pmatrix}%
\end{equation}%
Now considering $B^{\prime }=0$, and subsequently$B^{\prime }=1,$ in Eq.
(6a) we obtain 
\begin{equation}
l^{\mu }\left( \partial _{\mu }+iqA_{\mu }\right) P^{0}+\overline{m}^{\mu
}\left( \partial _{\mu }+iqA_{\mu }\right) P^{1}+\left( \Gamma
_{1000^{\prime }}-\Gamma _{0010^{\prime }}\right) P^{0}
\end{equation}%
\begin{equation*}
+\left( \Gamma _{1100^{\prime }}-\Gamma _{0110^{\prime }}\right) P^{1}-i\mu
_{0}\overline{Q}^{1^{\prime }}=0,
\end{equation*}%
\begin{equation}
m^{\mu }\left( \partial _{\mu }+iqA_{\mu }\right) P^{0}+n^{\mu }\left(
\partial _{\mu }+iqA_{\mu }\right) P^{1}+\left( \Gamma _{1001^{\prime
}}-\Gamma _{0011^{\prime }}\right) P^{0}
\end{equation}%
\begin{equation*}
+\left( \Gamma _{1101^{\prime }}-\Gamma _{0111^{\prime }}\right) P^{1}+i\mu
_{0}\overline{Q}^{0^{\prime }}=0,
\end{equation*}%
where $\mu _{0}=\sqrt{2}\mu _{p}$ and $q$ are the mass and the charge of the
Dirac particle respectively. $A_{\mu }$ is the electromagnetic vector
potential. Now by taking the complex conjugation of Eq. (6b) and writing
various spin coefficients by their named symbols as in [19,20] \ and
choosing $P^{0}=F_{1,\,}P^{1}=F_{2,\,}\overline{Q}^{0^{\prime }}=G_{2,\,}%
\overline{Q}^{1^{\prime }}=G_{1,\,}$ we obtain the Dirac equation in the NP
formalism [19,20] as 
\begin{subequations}
\begin{equation}
\left( l^{\mu }\partial _{\mu }+iql^{\mu }A_{\mu }+\epsilon -\rho \right)
F_{1}+\left( \overline{m}^{\mu }\partial _{\mu }+iq\overline{m}^{\mu }A_{\mu
}+\pi -\alpha \right) F_{2}=i\mu _{0}G_{1},
\end{equation}%
\begin{equation}
\left( n^{\mu }\partial _{\mu }+iqn^{\mu }A_{\mu }+\mu -\gamma \right)
F_{2}+\left( m^{\mu }\partial _{\mu }+iqm^{\mu }A_{\mu }+\beta -\tau \right)
F_{1}=i\mu _{0}G_{2},
\end{equation}%
\begin{equation}
\left( l^{\mu }\partial _{\mu }+iql^{\mu }A_{\mu }+\overline{\epsilon }-%
\overline{\rho }\right) G_{2}-\left( m^{\mu }\partial _{\mu }+iqm^{\mu
}A_{\mu }+\overline{\pi }-\overline{\alpha }\right) G_{1}=i\mu _{0}F_{2},
\end{equation}%
\begin{equation}
\left( n^{\mu }\partial _{\mu }+iqn^{\mu }A_{\mu }+\overline{\mu }-\overline{%
\gamma }\right) G_{1}-\left( \overline{m}^{\mu }\partial _{\mu }+iq\overline{%
m}^{\mu }A_{\mu }+\overline{\beta }-\overline{\overline{\tau }}\right)
G_{2}=i\mu _{0}F_{1}.
\end{equation}%
We will now study the Dirac equation (10) in the background of metric (1).
Let us write the null tetrad basis vectors of the null tetrad as \bigskip 
\end{subequations}
\begin{equation*}
l_{\mu }=dt-\frac{r^{2}dr}{\sqrt{\Delta }},
\end{equation*}%
\begin{equation*}
n_{\mu }=\frac{\Delta }{2r^{2}}dt+\frac{1}{2}dr,
\end{equation*}%
\ 
\begin{equation}
m_{\mu }=\frac{-r}{\sqrt{2}}(d\theta +i\sin \theta d\phi ),
\end{equation}%
and 
\begin{equation*}
l^{\mu }=\frac{r^{2}}{\Delta }dt+dr,
\end{equation*}%
\begin{equation*}
n^{\mu }=\frac{1}{2}dt-\frac{\Delta }{2r^{2}}dr,
\end{equation*}%
\ 
\begin{equation}
m^{\mu }=\frac{1}{\sqrt{2}r}(d\theta +\frac{i}{\sin \theta }d\phi ),
\end{equation}

Using the above tetrad we determine the nonzero NP complex spin coefficients
[20] as, 
\begin{equation*}
\rho =-\frac{1}{r},\qquad \mu =\frac{-\Delta }{2r^{3}},
\end{equation*}

\begin{equation*}
\gamma =\frac{r-M}{2r^{2}}-\frac{\Delta }{2r^{3}},
\end{equation*}

\begin{equation}
\alpha =-\beta =\frac{-\cot \theta }{2\sqrt{2}r}.
\end{equation}

We will consider the corresponding Compton wave of the Dirac particle as in
the form of $F=F\left( r,\theta \right) e^{i\left( kt+m\phi \right) }$ ,
where $k$ is the frequency of the incoming wave and $m$ is the azimuthal
quantum number of the wave. The form of the Dirac equation suggests that we
assume [19], 
\begin{equation}
rF_{1}=f_{1}\left( r,\theta \right) e^{i\left( kt+m\phi \right) },  \notag
\end{equation}%
\begin{equation*}
F_{2}=f_{2}\left( r,\theta \right) e^{i\left( kt+m\phi \right) },
\end{equation*}%
\begin{equation*}
G_{1}=g_{1}\left( r,\theta \right) e^{i\left( kt+m\phi \right) },
\end{equation*}%
\begin{equation}
rG_{2}=g_{2}\left( r,\theta \right) e^{i\left( kt+m\phi \right) }.
\end{equation}

\bigskip Substituting the appropriate spin coefficients (13) and the spinors
(14) into the Dirac equation (10) , we obtain

\begin{subequations}
\begin{equation}
\mathbf{D}f_{1}+\frac{1}{\sqrt{2}}\mathbf{L}f_{2}=i\mu _{0}rg_{1},
\end{equation}%
\begin{equation}
\frac{\Delta }{2}\mathbf{D}^{\dag }f_{2}-\frac{1}{\sqrt{2}}\mathbf{L}%
^{^{\dag }}f_{1}=-i\mu _{0}rg_{2},
\end{equation}%
\begin{equation}
\mathbf{D}g_{2}-\frac{1}{\sqrt{2}}\mathbf{L}^{^{\dag }}g_{1}=i\mu _{0}rf_{2},
\end{equation}%
\begin{equation}
\frac{\Delta }{2}\mathbf{D}^{\dag }g_{1}+\frac{1}{\sqrt{2}}\mathbf{L}%
g_{2}=-i\mu _{0}rf_{1},
\end{equation}%
where 
\end{subequations}
\begin{subequations}
\begin{equation}
\mathbf{D}=\frac{d}{dr}+i\frac{kr^{2}}{\Delta }+i\frac{qMr^{2}\sqrt{1-a^{2}}%
}{\Delta 2a\left( r-r_{h}\right) }
\end{equation}%
\begin{equation}
\mathbf{D}^{\dag }=\frac{d}{dr}-i\frac{kr^{2}}{\Delta }-i\frac{qMr^{2}\sqrt{%
1-a^{2}}}{\Delta 2a\left( r-r_{h}\right) }+\frac{r-M}{\Delta }
\end{equation}%
\begin{equation}
\mathbf{L}=\frac{d}{d\theta }+\frac{m}{\sin \theta }+\frac{\cot \theta }{2}
\end{equation}%
\begin{equation}
\mathbf{L}^{^{\dag }}=\frac{d}{d\theta }-\frac{m}{\sin \theta }+\frac{\cot
\theta }{2}
\end{equation}%
It is now apparent that Eqs. (15) can be separated by implying the
separability ansatz 
\end{subequations}
\begin{eqnarray}
f_{1} &=&R_{1}\left( r\right) A_{1}\left( \theta \right) ,\qquad
f_{2}=R_{2}\left( r\right) A_{2}\left( \theta \right) , \\
g_{1} &=&R_{2}\left( r\right) A_{1}\left( \theta \right) ,\qquad
g_{2}=R_{1}\left( r\right) A_{2}\left( \theta \right) .  \notag
\end{eqnarray}%
With this ansatz, Eqs. (15) become 
\begin{subequations}
\begin{equation}
A_{1}\mathbf{D}R_{1}+\frac{1}{\sqrt{2}}R_{2}\mathbf{L}A_{2}=i\mu
_{0}rR_{2}A_{1},
\end{equation}%
\begin{equation}
\Delta A_{2}\mathbf{D}^{\dag }R_{2}-\sqrt{2}R_{1}\mathbf{L}^{^{\dag
}}A_{1}=-2i\mu _{0}rR_{1}A_{2},
\end{equation}%
\begin{equation}
A_{2}\mathbf{D}R_{1}-\frac{1}{\sqrt{2}}R_{2}\mathbf{L}^{^{\dag }}A_{1}=i\mu
_{0}rR_{2}A_{2},
\end{equation}%
\begin{equation}
\Delta A_{1}\mathbf{D}^{\dag }R_{2}+\sqrt{2}R_{1}\mathbf{L}A_{2}=-2i\mu
_{0}rR_{1}A_{1}.
\end{equation}%
These equations (18) imply that 
\end{subequations}
\begin{subequations}
\begin{eqnarray}
\mathbf{D}R_{1}-i\mu _{0}rR_{2} &=&-\lambda _{1}R_{2},\qquad \\
\Delta \mathbf{D}^{\dag }R_{2}+2i\mu _{0}rR_{1} &=&\lambda _{2}R_{1}, \\
\mathbf{D}R_{1}-i\mu _{0}rR_{2} &=&\lambda _{3}R_{2}, \\
\Delta \mathbf{D}^{\dag }R_{2}+2i\mu _{0}rR_{1} &=&-\lambda _{4}R_{1},
\end{eqnarray}%
\begin{eqnarray}
\mathbf{L}A_{2} &=&\lambda _{1}A_{1},\qquad \mathbf{L}^{^{\dag
}}A_{1}=\lambda _{2}A_{2},\qquad \\
\mathbf{L}^{^{\dag }}A_{1} &=&\lambda _{3}A_{2},\qquad \mathbf{L}%
A_{2}=\lambda _{4}A_{1},
\end{eqnarray}%
where $\lambda _{1},$ $\lambda _{2},$ $\lambda _{3}$ and $\lambda _{4}$ are
four constants of separation. However, let us assume $\left( \lambda
_{4}=\lambda _{1}=-\lambda ,\lambda _{2}=\lambda _{3}=\lambda \right) ,$then
we obtain the radial and the angular pair equations

\end{subequations}
\begin{equation}
\mathbf{D}R_{1}=\left( \lambda +i\mu _{0}r\right) R_{2},\qquad \Delta 
\mathbf{D}^{\dag }R_{2}=\left( \lambda -2i\mu _{0}r\right) R_{1},
\end{equation}%
\begin{equation}
\mathbf{L}A_{2}=-\lambda A_{1},\qquad \mathbf{L}^{^{\dag }}A_{1}=\lambda
A_{2}.
\end{equation}

\section{Solution of angular and Radial equations}

$\qquad $Angular Eqs. (21) can be written as

\begin{equation}
\frac{dA_{1}}{d\theta }+\left( \frac{\cot \theta }{2}-\frac{m}{\sin \theta }%
\right) A_{1}=-\lambda A_{2},
\end{equation}%
\begin{equation}
\frac{dA_{2}}{d\theta }+\left( \frac{\cot \theta }{2}+\frac{m}{\sin \theta }%
\right) A_{2}=\lambda A_{1}.
\end{equation}%
which are exactly the same as the angular equation in the S geometry whose
solution is given in terms of standard spherical harmonics [21-23] as

\begin{equation}
A_{1,2}=Y_{l}^{n}\left( \theta \right) ,
\end{equation}

with $\lambda ^{2}=\left( n+\frac{1}{2}\right) ^{2}$.

The redial equations (20) can be rearranged as%
\begin{equation}
\sqrt{\Delta }\left( \frac{d}{dr}+i\frac{kr^{2}}{\Delta }+i\frac{qMr^{2}%
\sqrt{1-a^{2}}/2a}{\Delta \left( r-r_{h}\right) }\right) R_{1}=\left(
\lambda +i\mu _{\ast }r\right) \sqrt{\Delta }R_{2},
\end{equation}%
\begin{equation}
\sqrt{\Delta }\left( \frac{d}{dr}-i\frac{kr^{2}}{\Delta }-i\frac{qMr^{2}%
\sqrt{1-a^{2}}/2a}{\Delta \left( r-r_{h}\right) }\right) \sqrt{\Delta }%
R_{2}=\left( \lambda -i\mu _{\ast }r\right) R_{1},
\end{equation}%
where $\mu _{\ast }$ is the normalized rest mass of the spin-1/2 particle.

Our task now is to put the radial equations (25) and (26) in the form of one
dimensional wave equations. Therefore, to achieve our task we follow the
method applied by Chandrasekhar's book [19]. Starts with the transformations

\begin{equation}
P_{1}=R_{1},\qquad P_{2}=\sqrt{\Delta }R_{2}.
\end{equation}%
With these transformation Eqs. (25) and (26) become%
\begin{equation}
\frac{dP_{1}}{dr}+i\frac{k\Omega }{\Delta }P_{1}=\frac{1}{\Delta }\left(
i\mu _{\ast }r+\lambda \right) P_{2},
\end{equation}%
\begin{equation}
\frac{dP_{2}}{dr}-i\frac{k\Omega }{\Delta }P_{2}=\frac{1}{\Delta }\left(
\lambda -i\mu _{\ast }r\right) P_{1},
\end{equation}%
where 
\begin{equation}
\Omega =r^{2}+\frac{qMr^{2}\sqrt{1-a^{2}}}{2a\left( r-r_{h}\right) k}
\end{equation}%
Assume 
\begin{equation}
\frac{du}{dr}=\frac{\Omega }{\Delta }.
\end{equation}%
Then, the above equations (28) and (29) in terms of the new independent
variable $u,$ become%
\begin{equation}
\frac{dP_{1}}{du}+ikP_{1}=\frac{\sqrt{\Delta }}{\Omega }\left( i\mu _{\ast
}r+\lambda \right) P_{2},\qquad
\end{equation}%
\begin{equation}
\frac{dP_{2}}{du}-ikP_{2}=\frac{\sqrt{\Delta }}{\Omega }\left( \lambda -i\mu
_{\ast }r\right) P_{1}.
\end{equation}%
where 
\begin{eqnarray}
u &=&r-\sqrt{C}\tan ^{-1}\left( \frac{r}{\sqrt{C}}\right) -\frac{qMr^{2}%
\sqrt{1-a^{2}}}{2ak\left( 2r_{h}^{2}+2C\right) }[\frac{2r_{h}}{\sqrt{C}}\tan
^{-1}\left( \frac{r}{\sqrt{C}}\right)  \notag \\
&&-2\log \left( r-r_{h}\right) +\log \left( C+r^{2}\right) ],
\end{eqnarray}%
and $C=M^{2}-a^{2}M^{2}-2Mr.$

Let us apply another transformation, namely the new functions 
\begin{equation}
P_{1}=\phi _{1}\exp \left( \frac{-i}{2}\tan ^{-1}\left( \frac{\mu _{\ast }r}{%
\lambda }\right) \right) ,\qquad P_{2}=\phi _{2}\exp \left( \frac{i}{2}\tan
^{-1}\left( \frac{\mu _{\ast }r}{\lambda }\right) \right) ,
\end{equation}%
and next changing the variable $u$ into $\widehat{r}$ as $\widehat{r}=u-%
\frac{1}{2k}\tan ^{-1}\left( \frac{\mu _{0}r}{\lambda }\right) ,$ then we
can write Eqs.(32) and (33) in the alternative forms 
\begin{equation}
\frac{d\phi _{1}}{d\widehat{r}}+ik\phi _{1}=W\phi _{2},
\end{equation}%
\begin{equation}
\frac{d\phi _{2}}{d\widehat{r}}-ik\phi _{2}=W\phi _{1},
\end{equation}%
where 
\begin{equation}
W=\frac{2k\sqrt{\Delta }\left( \lambda ^{2}+\mu _{\ast }^{2}r^{2}\right)
^{3/2}}{2kr^{2}\left( \lambda ^{2}+\mu _{\ast }^{2}r^{2}\right) \left( 1+%
\frac{qM\sqrt{1-a^{2}}}{2a\left( r-r_{h}\right) k}\right) +\Delta \lambda
\mu _{\ast }}.
\end{equation}

Finally, in order to put the above equations (36) and (37) into one
dimensional wave equations, we define%
\begin{equation}
2\phi _{1}=\psi _{1}+\psi _{2},\qquad 2\phi _{2}=\psi _{1}-\psi _{2}.
\end{equation}%
Then Eqs.(36) and (37) become\bigskip 
\begin{equation}
\frac{d\psi _{1}}{d\widehat{r}}-W\psi _{1}=-ik\psi _{2},
\end{equation}%
\begin{equation}
\frac{d\psi _{2}}{d\widehat{r}}+W\psi _{2}=-ik\psi _{1}.
\end{equation}%
Which can be cast into 
\begin{equation}
\frac{d^{2}\psi _{1}}{d\widehat{r}^{2}}+k^{2}\psi _{1}=V_{+}\psi _{1},
\end{equation}%
\begin{equation}
\frac{d^{2}\psi _{2}}{d\widehat{r}^{2}}+k^{2}\psi _{2}=V_{-}\psi _{2},
\end{equation}%
where the effective potentials can be obtained from 
\begin{equation}
V_{\pm }=W^{2}\pm \frac{dW}{d\widehat{r}}.
\end{equation}%
We calculate the potentials as%
\begin{eqnarray}
V_{\pm } &=&\frac{\Delta B^{3}}{D^{2}}\pm \frac{\sqrt{\Delta }B^{3/2}}{D^{2}}%
\left( \left( r-M\right) B+3\Delta r\mu _{\ast }^{2}\right) \\
&&\mp \frac{\Delta ^{3/2}B^{5/2}}{D^{3}}\left[ \left( 2rB+2r^{3}\mu _{\ast
}^{2}\right) \left( 1+\frac{qM\sqrt{1-a^{2}}}{2a\left( r-r_{h}\right) k}%
\right) -r^{2}B\frac{qM\sqrt{1-a^{2}}}{2a\left( r-r_{h}\right) ^{2}k}+\frac{%
\left( r-M\right) \lambda \mu _{\ast }}{k}\right] ,  \notag
\end{eqnarray}%
where 
\begin{equation}
B=\left( \lambda ^{2}+\mu _{\ast }^{2}r^{2}\right) ,\qquad D=r^{2}B\left( 1+%
\frac{qM\sqrt{1-a^{2}}}{2a\left( r-r_{h}\right) k}\right) +\frac{\Delta
\lambda \mu _{\ast }}{2k}.
\end{equation}

Let us note that for the case of $a=1$, the potentials reduced to the the S
geometry, namely%
\begin{eqnarray}
V_{\pm } &=&\frac{\Delta L^{3}}{\left( r^{2}B+\frac{\Delta \lambda \mu
_{\ast }}{2k}\right) ^{2}}\pm \frac{\sqrt{\Delta }B^{3/2}}{\left( r^{2}B+%
\frac{\Delta \lambda \mu _{\ast }}{2k}\right) ^{2}}\left( \left( r-M\right)
B+3\Delta r\mu _{\ast }^{2}\right)  \\
&&\mp \frac{\Delta ^{3/2}B^{5/2}}{\left( r^{2}B+\frac{\Delta \lambda \mu
_{\ast }}{2k}\right) ^{3}}\left[ 2rB+2r^{3}\mu _{\ast }^{2}+\frac{\left(
r-M\right) \lambda \mu _{\ast }}{k}\right] .  \notag
\end{eqnarray}%
Here we are going to find the complete solution of Eqs. (42) and (43). First
we can rewrite the equations as%
\begin{equation}
\frac{d^{2}\psi _{1}}{d\widehat{r}^{2}}+\left( k^{2}-V_{+}\right) \psi
_{1}=0,
\end{equation}%
\begin{equation}
\frac{d^{2}\psi _{2}}{d\widehat{r}^{2}}+\left( k^{2}-V_{-}\right) \psi
_{2}=0.
\end{equation}%
This is simply the one dimensional Schr\"{o}dinger wave equations with
potentials $V_{\pm }$ and the total energy of the wave $k^{2}$. We can solve
Eqs. (48) and (49) by WKB\ approximation method [24,25]. The solution is
given by 
\begin{equation*}
\psi _{1}=\sqrt{T_{1}\left[ \omega _{1}\left( \widehat{r}\right) \right] }%
e^{iy_{1}}+\sqrt{R_{1}\left[ \omega _{1}\left( \widehat{r}\right) \right] }%
e^{-iy_{1}},
\end{equation*}%
\begin{equation}
\psi _{2}=\sqrt{T_{2}\left[ \omega _{2}\left( \widehat{r}\right) \right] }%
e^{iy_{2}}+\sqrt{R_{2}\left[ \omega _{2}\left( \widehat{r}\right) \right] }%
e^{-iy_{2}},
\end{equation}%
where 
\begin{equation}
\omega _{1}\left( \widehat{r}\right) =\sqrt{\left( k^{2}-V_{+}\right) }%
,\qquad \omega _{2}\left( \widehat{r}\right) =\sqrt{\left(
k^{2}-V_{-}\right) }
\end{equation}%
\begin{equation}
y_{1}\left( \widehat{r}\right) =\int \omega _{1}\left( \widehat{r}\right) d%
\widehat{r}+\text{constant},\qquad y_{2}\left( \widehat{r}\right) =\int
\omega _{2}\left( \widehat{r}\right) d\widehat{r}+\text{constant}
\end{equation}%
with 
\begin{equation}
T_{1}\left( r\right) +R_{1}\left( r\right) =1,\qquad T_{2}\left( r\right)
+R_{2}\left( r\right) =1\qquad \text{instantaneously.}
\end{equation}%
Where $\omega $ is the wave number of the incoming wave and $y$ is the 
\textit{eikonal. }$T_{1,2}$ and $R_{1,2}$ are the instantaneous transmission
and reflection coefficients [26] respectively. Let us note that the above
solution is valid when $\left( 1/\omega \right) \left( d\omega /d\widehat{r}%
\right) \ll \omega $. \bigskip 

\section{Discussion}

In this section, we are going to expose the effect of the external parameter 
$a$ on the effective potentials by plotting the potentials as a function of
radial distance with varying $a$. Next, we will study the behavior of the
potentials by drawing potentials curves for some different value of
frequency $k$ and charge $q$ of the spin-1/2 particle. From the potentials
Eq. (45), it is very clear that the potentials depend strictly on the
external parameter $a$ and the charge of the particle. Recall that the
external parameter $a$ has been discussed in details [7], indeed $a$
interpolates between two well-Known solutions, namely, the RN and BR
solutions. The physical result from this merging it that for a marginally
formed RN black hole the black hole property will be lost if the collapse
star radius $r_{s}$ satisfies $M^{2}+a\sqrt{M^{2}-e^{2}}<r_{s}<\sqrt{%
M^{2}-e^{2}}$, where $e$ is the black hole charge. For the case of S
embedded with BR solution, horizon shrinks as in (4).

To examine the effect of external parameter $a$ on the potentials we obtain
two dimensional plots of Eq. (45). In fig. 1, we exhibit the behaviour of
the potentials $V_{\pm }$ for different values of $a,$ the case $\left(
a=0\right) $ is excluded because it lead us to extramal RN case in the
metric, where we have chosen the rest mass to be $\mu _{\ast }=0.12,$ and
fixed values of $k=0.2,q=1$ such that $\mu _{\ast }<k$. It is seen from
fig.1 that, potentials have sharp peaks in the physical distance of $r,$ and
when the external parameter $a$ increases the level of the sharp peaks
increases. We conclude that for large value of $a,$ a massive charged
spin-1/2 particle moving in the physical region faces a high potential
barriers whereas for small value of $a$ encounters low potential barriers.
Therefore, as the external parameter increases the kinetic energy of the
particle decreases and hardly advances because of the strong retarding
potentials.

To examine the behaviour of the potentials for some values of frequencies we
obtain figure 2 for various value of $k$ and fixed value of $a=0.4$. It is
seen from fig. 2 that, the general behavior of \ the potentials are not
changed, they still have sharp peaks and behave similar for large distances.
Indeed, we notice that at high frequencies the sharp peaks are clear, but
for low frequencies the level of sharp peaks decrease. However, Figure 3
shows the nature of the potential changes for different values of the
particle charge $q$, where the fixed values are $\left( a=0.4\text{ and }%
k=0.2\right) .$ It is very clear from fig. 3 that, for small value of $q$,
the corresponding potential barrier have sharp peaks. It is also noticed
that as $q$ increases the level of sharp peaks decreases. The peaks seem to
disappear after a certain value of $q$. Therefore we conclude that
increasing the charge of the particle reduces the potential barriers and
resulting in increasing the kinetic energy of the particle.

Now, the three dimensional plots of the potentials with respect to the
external parameter and the radial distance $r$ is given in figures 4 and 5.
Figures 4 and 5 show the effect of the external parameter on the the
potentials explicitly for the massive charged spin-1/2 particle, for fixed
values of $\mu _{\ast }=0.12,q=1$ and $k=0.2.$ It is seen from figures 4 and
5 that, high potential barriers are observed for large values of external
parameter whereas for small value the potential barriers decrease. Again for
large distances, potentials levels decrease and asymptote behaviour is
manifested. The three dimensional plot of potential with respect to
frequency $k$, for fixed values of $a=0.4,q=1$ and $\mu _{\ast }=0.12$, and
the radial distance is given in figure 6. We observe \ from Fig. 6 that,
sharp peaks are clear for high frequencies only. Finally, the three
dimensional plot of potential with varying $q$ is given in fig. 7. We can
observe from fig. 7 that as the charge of the particle increases the
potential levels off.

\section{Conclusion}

In this paper, we have considered the Dirac equation in Schwarzschild black
hole immersed in an electromagnetic universe, using NP\ null tetrad
formalism. By employing an axially symmetric ansatz for the Dirac spinor, we
managed to separate the equation into radial and angular parts. The angular
equations obtained in this geometry are similar to the Schwarzschild case
where their solutions are given in terms of standard spherical harmonics.
For the radial equations we were able to obtain the radial wave equations
with effective potentials. Finally, we studied the behaviour of the
potentials by varying the external parameter, charge and the frequency of
the spin-1/2 particle, by plotting two and three dimensional plots. We
showed that, as the external parameter\ and the frequency increase,
potentials barriers become high and sharp peaks are clear. However, as the
charge of the particle increases potential levels decreases. Our main
motivation in the present paper paves the way to study the quasi-normal
modes associated to a field of spin-1/2 on the SEBH background. Also the
given analytical expressions of the solution could be useful for further
study of the thermodynamical properties of the spinor field in same
background. For future work, we may generalized our present case by studying
the Dirac particles in RN\ black hole coupled to an external, stationary
electromagnetic field.

\bigskip

\bigskip

{\LARGE References}

\bigskip

[1] Halilsoy M 1993 Gen. Relativ. Gravit. \textbf{25} 275.

[2] Halilsoy M 1993 Gen. Relativ. Gravit. \textbf{25} 975.

[3] George A Alekseev and Alberto A Garcia 1996 Phys. Rev. D \textbf{53},
1853.

[4] Robinson I 1959 Bull. Acad. Pol. Sci. \textbf{7},351.

[5] Bertotti B 1959 Phys. Rev. \textbf{116} 1331.[4] Halilsoy M and
Al-Badawi A1995 Class. Quantum Grav. \textbf{12} 3013.

[6] Halilsoy M and Al-Badawi A1995 Class. Quantum Grav. \textbf{12} 3013.

[7] Halilsoy M and Al-Badawi A 1998 IL Nuovo Cimento B \textbf{113} 761.

[8] Jin W M 1998 Class. Quantum Grav. \textbf{15} 3163.

[9] Semiz I 1992 Phys. Rev. D \textbf{46}, 5414.

[10] Mukhopadhyay B 1999 Ind. J. Phys. B \textbf{73 }855.

[11] Mukhopadhyay B 2000 Class. Quantum Grav. \textbf{17} 2017.

[12] Antonio Zecca 2006 Int. J. Theor. Phys. \textbf{45} 12.

[13] Antonio Zecca 2007 IL Nuovo Cimento B \textbf{121 }943.

[14] Cebeci H and \H{O}zdemir N 2013 Class. Quantum Grav. \textbf{30} 175005.

[15] Birkandan T and Horta\c{c}su M 2007 J. Math. Phys. \textbf{48} 092301.

[16] Belgiorno F and Cacciatori S L 2010 J. Math. Phys. \textbf{51} 033517.

[17] Finster F, Smoller J and Yau S T 2000 J. Math. Phys. \textbf{41} 2173.

[18] Winklmeier M and Yamadan O 2006 J. Math. Phys. \textbf{47} 102503.

[19] Chandrasekhar S 1983 The Mathematical Theory of Black Holes Clarendon,
London

[20] Newman E Tand Penrose R 1962 J. Math. Phys. \textbf{3} 566.

[21]\ Chakrabarti S K 1984 Proc. R. Soc. A \textbf{391 }27.

[22] Newman E and Penrose R 1966 J. Math. Phys. \textbf{7} 863.

[23]\ Goldberg J N, Macfarlane A J, Newman E T, Rohrlich F and Sudarsan E C
G 1967 J. Math. Phys. \textbf{8} 2155.

[24] Davydov A M 1976 Quantum Mechanics 2nd edn (Oxford: Pergamon).

[25] Mathews J andWalker R L 1970 Mathematical Methods Of Physics 2nd edn
(New York: Benjamin-Cummings).

[26] Mukhopadhyay B and Chakrabarti S K 1999 Class. Quantum Grav. \textbf{16}
3165.

\bigskip

\FRAME{ftbpFU}{3.9539in}{2.4509in}{0pt}{\Qcb{Family of potentials graphs $%
V_{-}$( solid curves) and $V_{+}$( dashed curves) for different values of
external parameter a, with $%
%TCIMACRO{\U{3bc} }%
%BeginExpansion
\protect\mu
%EndExpansion
\ast =0.12,k=0.2,M=0.5$ and $q=1=\protect\lambda $. From the upper to the
lower curves the external parameter a is chosen as $0.8,0.6,0.4,0.2.$}}{}{%
Figure 1.}{\special{language "Scientific Word";type
"GRAPHIC";maintain-aspect-ratio TRUE;display "USEDEF";valid_file "T";width
3.9539in;height 2.4509in;depth 0pt;original-width 6.0416in;original-height
3.7282in;cropleft "0";croptop "1";cropright "1";cropbottom "0";tempfilename
'OKUIYU00.wmf';tempfile-properties "XPR";}}

\bigskip

\bigskip \FRAME{ftbpFU}{3.9539in}{2.4509in}{0pt}{\Qcb{Family of potentials
graphs $V_{-}$( solid curves) and $V_{+}$( dashed curves) for different
values of frequency $k$, with $%
%TCIMACRO{\U{3bc} }%
%BeginExpansion
\protect\mu
%EndExpansion
\ast $ $=0.12,a=0.4,M=0.5$ and $q=1=\protect\lambda $. From the upper to the
lower curves the k is chosen as $1,0.8,0.6,0.4,0.2.$}}{}{Figure 2.}{\special%
{language "Scientific Word";type "GRAPHIC";maintain-aspect-ratio
TRUE;display "USEDEF";valid_file "T";width 3.9539in;height 2.4509in;depth
0pt;original-width 6.0416in;original-height 3.7282in;cropleft "0";croptop
"1";cropright "1";cropbottom "0";tempfilename
'OKUJ9E01.wmf';tempfile-properties "XPR";}}

\bigskip

\FRAME{ftbpFU}{3.9539in}{2.4509in}{0pt}{\Qcb{ Family of potentials graphs $%
V_{-}$( solid curves) and $V_{+}$( dashed curves) for different values of
charge $q$, with $%
%TCIMACRO{\U{3bc} }%
%BeginExpansion
\protect\mu
%EndExpansion
\ast =0.12,k=0.2,M=0.5,a=0.4$ and $\protect\lambda =1$. From the lower to
the upper curves the charge q is chosen as $1,0.8,0.6,0.4,0.2.$}}{}{Figure
3. }{\special{language "Scientific Word";type
"GRAPHIC";maintain-aspect-ratio TRUE;display "USEDEF";valid_file "T";width
3.9539in;height 2.4509in;depth 0pt;original-width 6.0416in;original-height
3.7282in;cropleft "0";croptop "1";cropright "1";cropbottom "0";tempfilename
'OKUJD202.wmf';tempfile-properties "XPR";}}

\bigskip

\FRAME{ftbpFU}{3.9539in}{3.8709in}{0pt}{\Qcb{Three dimensional plot of the
potential $V_{-}$ for different values of external parameter$a$, with $%
%TCIMACRO{\U{3bc} }%
%BeginExpansion
\protect\mu
%EndExpansion
\ast =0.12,k=0.12,M=0.5$ and $q=1=\protect\lambda $.}}{}{Figure 4.}{\special%
{language "Scientific Word";type "GRAPHIC";maintain-aspect-ratio
TRUE;display "USEDEF";valid_file "T";width 3.9539in;height 3.8709in;depth
0pt;original-width 6.0416in;original-height 5.9127in;cropleft "0";croptop
"1";cropright "1";cropbottom "0";tempfilename
'OKUJGD03.wmf';tempfile-properties "XPR";}}

\FRAME{ftbpFU}{3.9539in}{3.8709in}{0pt}{\Qcb{Three dimensional plot of the
potential $V_{+}$ for different values of external parameter$a$, with $%
%TCIMACRO{\U{3bc} }%
%BeginExpansion
\protect\mu
%EndExpansion
\ast =0.12,k=0.12,M=0.5$ and $q=1=\protect\lambda .$}}{}{Figure 5.}{\special%
{language "Scientific Word";type "GRAPHIC";maintain-aspect-ratio
TRUE;display "USEDEF";valid_file "T";width 3.9539in;height 3.8709in;depth
0pt;original-width 6.0416in;original-height 5.9127in;cropleft "0";croptop
"1";cropright "1";cropbottom "0";tempfilename
'OKUJJ304.wmf';tempfile-properties "XPR";}}

\bigskip

\FRAME{ftbpFU}{3.9539in}{3.8709in}{0pt}{\Qcb{Three dimensional plot of the
potential $V_{+}$ for different values of frequency $k$, with $%
%TCIMACRO{\U{3bc} }%
%BeginExpansion
\protect\mu
%EndExpansion
\ast =0.12,a=0.4,M=0.5$ and $q=1=\protect\lambda $.}}{}{Figure 6. }{\special%
{language "Scientific Word";type "GRAPHIC";maintain-aspect-ratio
TRUE;display "USEDEF";valid_file "T";width 3.9539in;height 3.8709in;depth
0pt;original-width 6.0416in;original-height 5.9127in;cropleft "0";croptop
"1";cropright "1";cropbottom "0";tempfilename
'OKUJLM05.wmf';tempfile-properties "XPR";}}

\bigskip

\FRAME{ftbpFU}{3.9539in}{3.8709in}{0pt}{\Qcb{Three dimensional plot of the
potential $V_{-}$ for different values of charge $q$, with $%
%TCIMACRO{\U{3bc} }%
%BeginExpansion
\protect\mu
%EndExpansion
\ast =0.12,a=0.4,M=0.5,k=0.2$ and $\protect\lambda =1$.}}{}{Figure 7.}{%
\special{language "Scientific Word";type "GRAPHIC";maintain-aspect-ratio
TRUE;display "USEDEF";valid_file "T";width 3.9539in;height 3.8709in;depth
0pt;original-width 6.0416in;original-height 5.9127in;cropleft "0";croptop
"1";cropright "1";cropbottom "0";tempfilename
'OKUJOM06.wmf';tempfile-properties "XPR";}}

\end{document}